\documentclass[fleqn]{article}
\usepackage{ifthen}
\usepackage{amstext}
\usepackage{amssymb}
\usepackage{cite}

\title{Self-Organized Bottleneck\\ and Coexistence of
Incongruous States\\ in a Microwave Phonon Laser (Phaser)}

\author{
\textbf{
 D.~N.~Makovetskii
 }\\
  \\
Institute of Radiophysics and Electronics NASU,\\ 12, Academician
Proskura Str., Kharkov 61085, Ukraine.
\\
E-mail: \textit{dmakov@rambler.ru}
} 

\date{January 3, 2009}
\begin{document}

\maketitle

\begin{abstract}
{
%
Phenomena of emergence of regular and
chaotic fine structure (FS) in stimulated emission (SE) power
spectra of an autonomous microwave phonon laser (phaser) have been
revealed and investigated experimentally in pink ruby at liquid
helium temperatures. The phenomenon of a self-organized bottleneck
in evolution of the microwave acoustic FS lines has been observed
by means of narrow-range phonon SE spectral analysis. The
large-scale phenomenon of coexistence of incongruous (stationary,
periodic and chaotic) states in the whole spin-phonon phaser system
is revealed in experiments with panoramic power spectra of phonon
SE. Phaser active medium is characterized by a very small ratio $R
= T_f/T_a << 1$ (see D.~N.~Makovetskii, arXiv:cond-mat/0402640v1),
where $T_f$ is the lifetime of an emitted field quasiparticle
(namely, acoustic branch phonon), and $T_a$ is the time of
spin-lattice relaxation of an individual active center (AC). Such
the medium may be considered as an excitable system by analogy with
class-$B$ optical lasers (see C.~O.~Weiss e.a., Phys. Rev. A,
vol.47, p.R1616, 1993), if one takes into account local
(dipole-dipole) interactions between ACs. We propose a possible
direction for modeling of both the observed phenomena on the basis
of three-level cellular automata (S.~D.~Makovetskiy and
D.~N.~Makovetskii, arXiv:cond-mat/0410460v2; S.~D.~Makovetskiy,
arXiv:cond-mat/0602345v1) which emulate evolution of a bounded
phaser-like excitable system with locally interacting ACs in the
limit $R \rightarrow 0$.}
\end{abstract}

\clearpage

\section*{Introduction}\label{se:0}

The problem of studying the nonlinear processes occuring in open
dissipative systems which possess a complex phase-space structure
is challenging for a number of branches of modern physics
\cite{b01-Nicolis-1995}. In particular, the laser-based and other
nonlinear optical systems are studied intensively in this field of
science nowadays \cite{b02-Weiss-1999,b03-Weiss-2007}, and a number
of important fundamental and applied results has already been
obtained. The ideas of quantum electronics have also deeply
penetrated into adjacent branches of modern physics, especially
into quantum acoustics
\cite{b04-TuckerE-1961,b05-TuckerE-1964,b06-Shiren-1965,b07-TuckerE-1966,b08-Peterson-1969,b09-SSC-1974,b10-DAN-1975,b11-Golenishchev-1976,b12-JETP-1977,b13-JETPL-1978}.
A series of new nonlinear phenomena, which arise under the quantum
amplification or generation of tera-, hyper- or ultrasonic
oscillations in nonequilibrium (inverted) acoustic and
nanomechanical dissipative systems of various origins, have been
predicted theoretically and discovered experimentally during last
years
\cite{b14-TPL-2001,b15-cond-mat-0303188,b16-Fokker-1997,b17-Tilstra-2001,b18-Bargatin-PRL-2003,b19-Bargatin-cond-mat-0304605,b20-Tilstra-2007}.
The researches concerning the conditions, under which the formation
of inverted states and the purposeful control over nonlinear
processes in such systems (the acoustic analogs of lasers) become
possible, are of significant independent interest.

As has already been emphasized in works
\cite{b14-TPL-2001,b15-cond-mat-0303188,b21-TP-2004,b22-cond-mat-0402640},
if one confines the consideration to the quantum amplification and
generation of \textit{hypersound}, i.e. the acoustic waves within
the microwave range of frequencies, the following circumstances
invoke the greatest interest. A microwave phonon laser (phaser)
emits acoustic radiation with a wavelength of 1--3~$\mu$m, which
corresponds to the wavelength of electromagnetic radiation in the
near-infrared range, where many ordinary lasers generate as well.
At the same time, the velocity of hypersound in the phaser active
paramagnetic media is 5 orders of magnitude lower than the velocity
of electromagnetic waves. Accordingly, the frequency of stimulated
emission (SE) of acoustic waves in a paramagnetic phaser amounts to
$\omega=$~3 -- 10~GHz only, which is about the same 5 orders of
magnitude lower than the frequency of the electromagnetic SE in
optical lasers. Therefore, the relative intensity of the
spontaneous component $J_{\mathrm{spont}}$ of emission in a
microwave phaser is about 15 orders of magnitude lower than that in
an ordinary optical (electromagnetic) laser (because
$J_{\mathrm{spont}} \propto \omega^3$), whereas the lengths of
corresponding acoustic and electromagnetic waves coincide.

The low level of the quantum noise in phasers opens new
opportunities for studying the nonlinear phenomena in
nonequilibrium systems. This conclusion has already been confirmed
experimentally for a paramagnetic phaser with a pink ruby crystal
($\mathrm{{Cr}^{3+}} \: {:} \: \mathrm{{Al}_2{O}_3}$). In works
\cite{b14-TPL-2001,b15-cond-mat-0303188,b21-TP-2004,b22-cond-mat-0402640},
a new nonlinear resonance ($\lambda$-resonance), which is
accompanied by a number of interesting dynamical phenomena, has
been revealed experimentally in a non-autonomous ruby phaser and
studied in details at liquid- helium temperatures and
electromagnetic pumping with carrier frequency about 23~GHz. This
$\lambda$-resonance arises at modulation of pumping (or Zeeman
magnetic field) in the range of ultralow frequencies $\omega_m
\approx \omega_{\lambda}$, where $\omega_{\lambda} / 2 \pi \approx
10$~Hz
\cite{b14-TPL-2001,b15-cond-mat-0303188,b21-TP-2004,b22-cond-mat-0402640}.
It is considerably lower than the typical frequencies
$\omega_{\nu}$ of the usual relaxational resonance
(${\nu}$-resonance) in the ruby phaser ($\omega_{\nu} / 2 \pi
\approx$ 70--300~Hz), which was observed earlier in works
\cite{b23-TP-1989,b24-TP-1992} (note that ${\nu}$-resonance in the
ruby phaser is an acoustic analog of the well-known laser
relaxational resonance).

In works
\cite{b14-TPL-2001,b15-cond-mat-0303188,b21-TP-2004,b22-cond-mat-0402640},
it has been discovered experimentally that a plenty of coexisting
states of the spin-phonon system emerges in the domain of the
$\lambda$-resonance and the very slow regular transitions, which
remind autowave motions, occur between them. The period of the full
cycle of such transitions reaches giant values and exceeds $10^3 -
10^4$~s at $\omega_m \approx \omega_{\lambda}$
\cite{b14-TPL-2001,b15-cond-mat-0303188,b21-TP-2004,b22-cond-mat-0402640}.
This is several orders of magnitude longer than the period $\tau_m
\equiv 2\pi / \omega_m$ of an external force that modulates phaser
pumping (or Zeeman magnetic field).

It should be mentioned that a kind of slowing-down of transient
processes in paramagnetic crystals, which accompanies the
saturation of spin transitions, has already been studied earlier
\cite{b25-Jeffries-1963,b26-Abragam-1970,b27-Altshuler-1972} under
conditions of the so-called phonon bottleneck caused by
non-coherent (thermal) effects in a resonant spin-phonon system
which interacts with a thermostat. However, in the 1970-1980s, it
has been proved experimentally
\cite{b05-TuckerE-1964,b07-TuckerE-1966,b09-SSC-1974,b12-JETP-1977,b28-Diss-1983,b29-Aref-1984}
that, at least in the microwave range, both the quantum
amplification and self-excitation of phonons in phasers arise just
in the absence of the phonon bottleneck effect, which could only
inhibit the normal functioning of a phaser as a quantum oscillator.

Anyway, the phonon bottleneck effect in pink ruby in the microwave
range and at temperatures of liquid helium is absent both in our
system and in analogous paramagnetic systems used by other authors
in numerous investigations of quantum paramagnetic amplifiers. In
particular, our experimental data for pink ruby crystals
\cite{b09-SSC-1974,b12-JETP-1977,b28-Diss-1983,b29-Aref-1984} have
demonstrated that at temperatures $\theta = 1,7 - 4,2$~K the
maximal lifetime of resonant phonons with a frequency of about
10~GHz is about $10^4$ times shorter than the time of longitudinal
spin relaxation $\tau_1 \approx 0{,}1 - 1$~s of chromium ions. This
is the direct manifestation of nonoccurence of phonon bottleneck in
the active medium of our phaser.

Thus, the emergence of slow processes, which were observed in
paramagnetic non-autonomous phaser
\cite{b14-TPL-2001,b15-cond-mat-0303188,b21-TP-2004,b22-cond-mat-0402640},
cannot evidently be a result of influence of incoherent phonon
bottleneck. The slowing-down of the energy exchange in the active
medium of non-autonomous paramagneic phaser, as was shown in
\cite{b14-TPL-2001,b15-cond-mat-0303188,b21-TP-2004,b22-cond-mat-0402640},
reflects the peculiarities of the self-organization of dissipative
structures in the spin-phonon system of such a phaser, provided a
strong external perturbation ov the inversion states of
paramagnetic centers. Does such or similar slowing-down exists in
an \textit{autonomous} phaser, where the self-organization emerges
in absence of external forces at all? The main goal of this work
was experimental search for the answer to this question. The
results of theoretical researches and computer studies
\cite{b30-Morita-2004,b31-Morita-cond-mat-0304649,b32-Morita-cond-mat-0407460,b33-Awazu-2004,b34-Awazu-nlin-0310018,b35-SDM-cond-mat-0602345,b36-SDM-Master-Thesis-2006,b37-SDM-cond-mat-0410460},
which have been published recently, served as an additional
motivation for this work to be fulfilled.

\section{The Phaser System}\label{se:1}

The experimental part of the work was carried out making use of a
microwave phaser
\cite{b14-TPL-2001,b15-cond-mat-0303188,b21-TP-2004,b22-cond-mat-0402640}.
As it was already pointed out in works
\cite{b14-TPL-2001,b15-cond-mat-0303188}, a microwave phaser is an
acoustic analog of the class-$B$ optical lasers
\cite{b03-Weiss-2007}. In particular, both in class-$B$ lasers and
in a phaser the slow processes of the population inversion of
active centers (ACs) are the leading ones, while the fast-relaxing
system of the field excitations follows the AC-system evolution.

A phaser differs from a class-$B$ laser only by the nature of the
stimulated emission (SE). Namely, in a class-$B$ laser the SE is
purely electromagnetic, but in a phaser it is mainly acoustic. The
pointed difference is very essential, because velocity of acoustic
waves is typically 5 orders lower than velocity of light. We will
discuss this principal difference later.

The main element of the phaser, which was already described in
\cite{b14-TPL-2001,b15-cond-mat-0303188,b21-TP-2004,b22-cond-mat-0402640},
is a ruby acoustic Fabry--Perot resonator (AFPR) with two flat and
parallel acoustic mirrors. The AC-system of ruby phaser consists of
trivalent cromium (${\mathrm{Cr}}^{3+}$) paramagnetic ions in
non-magnetic corundum crystal matrix. The inversion of spin-levels
populations of AC-system in ruby phaser is formed by continuous
stationary pumping. The pumping field is a microwave
electromagnetic field excited from outside in an electromagnetic
pump cavity (EPC) with the AFPR inside.

At one of the acoustic mirrors, the thin-film piezoelectric
transducer (composed of a textured ${\mathrm{ZnO}}$ film and an
${\mathrm{Al}}$ interlayer) is mounted. This transducer admits to
detect an acoustic SE (laser-like phonon emission) that arises in
the ARFP when the corresponding (${\mathrm{Cr}}^{3+}$) spin
transitions are being pumped. In the simplest case the acoustic
microwave power spectrum of this SE consists of very narrow lines
located at eigenfrequencies of ARFP $\omega_n / 2 \pi
\approx$~9~GHz ($\Delta_n / 2 \pi \equiv (\omega_n - \omega_{n-1})
/ 2 \pi = 310$~kHz).

It is important that the thin-film piezoelectric transducer is
bi-directional. So, making use of a microwave external generator
(that excites microwave acoustic vibrations of the ${\mathrm{ZnO}}$
film), one can inject a phonon flux in the phaser active system
inependently from the laser-like phonon emission generated by ACs.
The frequency of this injected acoustic signal
$\omega_{\mathrm{inj}}$ is equal to the frequency of the microwave
external generator and generally is not equal to any of
eigenfrequencies of AFPR.

Other details concerning the thin-film piezoelectric transducer,
the AFPR, the EPC, and the control parameters (the static magnetic
field $\vec{H}$, the pump frequency $\omega_{\mathrm{pump}}$, the
pump power $P$, etc.), can be found in works
\cite{b21-TP-2004,b22-cond-mat-0402640}. However, in contrast to
works \cite{b21-TP-2004,b22-cond-mat-0402640}, all the measurements
of microwave phonon SE were carried out provided that any
modulation of the control parameters was absent. Therefore, all the
SE regimes observed in this work concern exclusively the autonomous
phaser.

\section{Absorption, Amplification and Self-Excitation of Hypersound
in the Phaser System}\label{se:2}

\subsection{Absorption of injected hypersound in the absence
of pumping}\label{subse:2-1}

Provided that the pumping is absent (the pump power $P = 0$), and
the amplitude and direction of $\vec{H}$ are far from the lines of
the acoustic paramagnetic resonance (APR) of the system
${\mathrm{Al}}_2{\mathrm{O}}_3 \: {:} \: {\mathrm{Cr}}^{3+}$, the
absorption of injected hypersound with the frequency
$\omega_{\mathrm{inj}}$, is governed by the two following
mechanisms:

(M1) Non-resonant attenuation of hypersound in the
${\mathrm{Al}}_2{\mathrm{O}}_3 \: {:} \: {\mathrm{Cr}}^{3+}$
crystal. This attenuation is characterized by the ``volume''
decrement $\eta_{\mathrm{vol}}$ of hypersound in the crystal medium
itself. The value $\eta_{\mathrm{vol}}$ depends, first of all, on
the crystal matrix perfection. In addition, this value includes
also hypersound energy leakage through the lateral (non-mirror)
surface of the AFPR;

(M2) Hypersound losses at the acoustic mirrors (one of which is
additively loaded by the thin-film piezoelectric transducer). This
attenuation is characterized by the effective ``mirror'' decrement
$\eta_{\mathrm{mirr}}$ of hypersound. The value
$\eta_{\mathrm{mirr}}$ takes into account leakage of the hypersound
energy through imperfect mirror surfaces of the AFPR. So it depends
strongly on the mirrors' quality and the transducer's loading.

Both mechanisms (M1) and (M2) lead to decreasing of intensity of
the injected hypersound in AFPR, because $\eta_{\mathrm{vol}}$ and
$\eta_{\mathrm{mirr}}$ are always positive.

If the injected hypersound frequency $\omega_{\mathrm{inj}}$ is
scanned, the acoustic microwave resonances are observed in the AFPR
at $\omega_{\mathrm{inj}} \approx \omega_n$. These acoustic
resonances are similar to ordinary electromagnetic ones in an
optical Fabry-Perot resonator. It is obvious that the pointed
acoustic resonances one can observe even in pure corundum
(${\mathrm{Al}}_2{\mathrm{O}}_3$) AFPR.

At $P = 0$ and $\omega_{\mathrm{inj}} = {\mathrm{const}}$, the
value and direction of $\vec{H}$ can be tuned in such a manner that
the APR arises in the ${\mathrm{Al}}_2{\mathrm{O}}_3 \: {:} \:
{\mathrm{Cr}}^{3+}$ paramagnetic system
\cite{b07-TuckerE-1966,b38-FTT-1973}. The APR is realized when the
splitting for a sertain pair of energy levels
${\mathcal{E}}_m(\vec{H})$ and ${\mathcal{E}}_n(\vec{H})$ in the
${\mathrm{Al}}_2{\mathrm{O}}_3 \: {:} \: {\mathrm{Cr}}^{3+}$ system
approximately coincides with $\hbar \omega_{\mathrm{inj}}$ and the
quantum transition ${\mathcal{E}}_m \leftrightarrow
{\mathcal{E}}_n$ is allowed for interaction with hypersound
\cite{b07-TuckerE-1966,b38-FTT-1973,b39-TuckerJ-1972}. Additively,
the intensity of injected hypersound $J_{\mathrm{inj}}$ must not be
excessively high, namely $J_{\mathrm{inj}} < J_{\mathrm{sat}}$,
where $J_{\mathrm{sat}}$ is saturation intensity for the quantum
transition ${\mathcal{E}}_m \leftrightarrow {\mathcal{E}}_n$ in
${\mathrm{Al}}_2{\mathrm{O}}_3 \: {:} \: {\mathrm{Cr}}^{3+}$
paramagnetic system.

Under these conditions, the additional losses of hypersound take
place at $\vec{H} \approx
\vec{H}_{\mathrm{inj}}^{(\mathrm{vert})}$, where
$\vec{H}_{\mathrm{inj}}^{(\mathrm{vert})}$ is the vertex of the
magnetic-field-line of the APR at the frequency
$\omega_{\mathrm{inj}}$. So, we have the third mechanism of the
absorption of injected hypersound in our system, namely:

(M3) Resonant paramagnetic attenuation of hypersound, which depends
strongly on $\omega_{\mathrm{inj}}$, $\vec{H}$ and
$J_{\mathrm{inj}}$, as well as on the hypersound type
(longitudinal, fast transverse, slow transverse)
\cite{b07-TuckerE-1966,b38-FTT-1973,b39-TuckerJ-1972}.

One can not observe this resonant paramagnetic attenuation in pure
corundum (${\mathrm{Al}}_2{\mathrm{O}}_3$) or in any other
non-magnetic medium. The (M3) kind of hypersound attenuation is
caused by spin-phonon interaction in paramagnetic medium (the ruby
crystal ${\mathrm{Al}}_2{\mathrm{O}}_3 \: {:} \:
{\mathrm{Cr}}^{3+}$ in our case).

Phenomenologically this attenuation is characterized by the
effective ``magnetic'' decrement
$\eta_{\mathrm{magn}}^{(\mathrm{null})}$ of hypersound. The
superscript $(\mathrm{null})$ indicates that pumping is absent ($P
= 0$). Quantitative description of the APR phenomenon and rigorous
formulae for hypersound attenuation one can find in
\cite{b07-TuckerE-1966,b38-FTT-1973,b39-TuckerJ-1972}

It is obvious, that $\eta_{\mathrm{magn}}^{(\mathrm{null})} > 0$
(as far as $\eta_{\mathrm{vol}}$ and $\eta_{\mathrm{mirr}}$), and
no amplification of injected hypersound in AFPR takes place in
absence of pumping.

\subsection{Quantum amplification of injected acoustic signal and
self-excitation of hypersound in ruby phaser}\label{subse:2-2}

Let the electromagnetic pumping is switched on ($P > 0$) at a
frequency $\omega_{\mathrm{pump}} = \omega_{\mathrm{cav}}$, where
$\omega_{\mathrm{cav}}$ is an eigenfrequency of the EPC. The vertex
$\vec{H}_{\mathrm{pump}}^{(\mathrm{vert})}$ of the
magnetic-field-line of the electron spin resonance (ESR) at the
frequency $\omega_{\mathrm{pump}}$ generally does not coincide with
that of the APR line $\vec{H}_{\mathrm{inj}}^{(\mathrm{vert})}$ at
the frequency $\omega_{\mathrm{inj}}$. But if the condition
$\vec{H}_{\mathrm{pump}}^{(\mathrm{vert})} =
\vec{H}_{\mathrm{inj}}^{(\mathrm{vert})}$ is satisfied (e.~g. after
an appropriate tuning of the frequency $\omega_{\mathrm{inj}}$),
and the ESR line is saturated, the APR line becomes inverted
\cite{b07-TuckerE-1966,b10-DAN-1975}. In other words, the
paramagnetic absorption of injected hypersound is changed by its
phaser amplification. As for the specific conditions for the
inversion of the APR line at the ESR saturation, see
\cite{b21-TP-2004,b22-cond-mat-0402640}.

If all the non-magnetic losses of hypersound ($\eta_{\mathrm{vol}}$
and $\eta_{\mathrm{mirr}}$) in the AFPR are totally compensated
owing to the phaser amplification, the self-excitation of
hypersound arises. This self-excitation is possible even in absence
of an injected signal: exponential growth of small noisy (thermal)
acoustic seed is the cause of the acoustic instability leading to
the hypersound self-excitation
\cite{b05-TuckerE-1964,b09-SSC-1974,b12-JETP-1977,b28-Diss-1983,b29-Aref-1984}.

Conditions of self-excitation of acoustic waves in a phaser are
much more comlicated than conditions of self-excitation of
electromagnetic waves in lasers. Two main causes of the pointed
complexity are as follows: (i) the coexistence of longitudinal,
transverse and mixed types of acoustic waves at an arbitrary
direction of hypersound propagation and (ii) very high anisotropy
of spin-phonon interaction in active paramagnetic crystals.

In this work, the geometrical axis of the AFPR
${\vec{\mathcal{O}}}_C$ coincides with the ruby's crystallographic
axis of the third order ${\vec{\mathcal{O}}}_3$. In this case no
mixed types of acoustic waves exist along the AFPR axis. Moreover,
it is known that, in the case of transverse acoustic waves with the
wave vector ${\vec{k}}_T \| {\vec{\mathcal{O}}}_3$, the so-called
conical refraction takes place \cite{b40-Truell-1972}. The last
phenomenon is a consequence of degeneration of fast and slow
transverse acoustic modes propagating along ${\vec{\mathcal{O}}}_3$
\cite{b40-Truell-1972}. Hence, only pure longitudinal hypersound
modes with the wave vector ${\vec{k}}_L \| {\vec{\mathcal{O}}}_C \|
{\vec{\mathcal{O}}}_3$ posses a high Q-factor in such an AFPR at
its microwave eigenfrequencies $\omega_n$.

When the pump power in our system is slowly increased beginning
from $P=0$, the considered compensation of non-magnetic losses of
hypersound achieves (step by step) for different logitudinal AFPR
modes $\omega_n$. First of all, self-excitation of hypersound
arises at the logitudinal AFPR mode $\omega_{1}$ which is located
most closely to the center of the inverted APR line $\omega_{0}$.
For this mode, the condition
\begin{equation}
  Q_{\mathrm{eff}}^{(1)} (P, \vec{H}, \omega_{1}) < 0,
  \label{eq:01}
\end{equation}
holds true ahead of all other modes $\omega_n$, as $P$ increases.
Here, $Q_{\mathrm{eff}}^{(1)}$ is the effective acoustic Q-factor
of the phaser system for the mode $\omega_1$ under pumping. This
effective acoustic Q-factor is determined by the relation:
\begin{equation}
  \frac{1}{Q_{\mathrm{eff}}^{(1)}} =
  \frac{1}{Q_{\mathrm{vol}}^{(1)}} +
  \frac{1}{Q_{\mathrm{mirr}}^{(1)}} +
  \frac{1}{Q_{\mathrm{magn}}^{(1)}}
  = \frac{1}{Q_C^{(1)}} + \frac{1}{Q_{\mathrm{magn}}^{(1)}},
  \label{eq:02}
\end{equation}
where $Q_{\mathrm{vol}}^{(1)} = k_L / \eta_{\mathrm{vol}}$;
$Q_{\mathrm{mirr}}^{(1)} = k_L / \eta_{\mathrm{mirr}}$; $k_L =
|{\vec{k}}_L| = \omega_{1} / V_L$; $V_L$ is the phase velocity of
hypersound; $Q_C^{(1)}$ the Q-factor of the AFPR at the $\omega_1$
mode in the absence of pumping, and $Q_{\mathrm{magn}}^{(1)}$ is
the magnetic Q-factor of this mode under pumping.

In contrast to $Q_C^{(1)}$, the magnetic $Q_{\mathrm{magn}}^{(1)}$
can be negative. It looks like
\begin{eqnarray}
  Q_{\mathrm{magn}}^{(1)}& = & - k_L / \alpha_{1} (P, \vec{H}, \omega_{1}) \equiv \nonumber \\
                         & \mbox{} \equiv & - k_L \left[ K(P, \vec{H}) \sigma(\vec{H}, \omega_{1}) \right]^{-1},
  \label{eq:03}
\end{eqnarray}
where $\alpha_{1}$ is the increment of quantum amplification of
hypersound for the mode under consideration; $K$ is the inversion
ratio for the APR line (in the case of inversion, $K > 0$ );
$\sigma \approx \eta_{\mathrm{magn}}^{(\mathrm{null})}$ is the
effective ``magnetic'' decrement of hypersound with frequency
$\omega_{1} \approx \omega_{0}$ in the absense of pumping.

Following works \cite{b21-TP-2004,b22-cond-mat-0402640}, we write
down the expression for $\sigma$ at the spin transition
${\mathcal{E}}_m \leftrightarrow {\mathcal{E}}_n$, in the form:
\begin{equation}
  \sigma_{mn} = \frac{ 2 {\pi}^2 C_a {\nu}^2 g(\nu) |\Phi_{mn}|^2  }%
  {(2S + 1) {\rho}^{\prime} V_L^3 k_B T },
  \label{eq:04}
\end{equation}
where $C_a$ is the concentration of paramagnetic centers;  $\nu =
\omega / 2 \pi$; $g(\nu)$ is the form-factor of the APR line;
${\rho}^{\prime}$ is the crystal density; $k_B$ is the Boltzmann
constant; $\Phi_{mn}$ is the coupling parameter of the spin
transition ${\mathcal{E}}_m \leftrightarrow {\mathcal{E}}_n$ with
hypersound.

The matrix element $\Phi_{mn}$ for a longitudinal hypersound wave
propagating along the ruby's axis of the third order
$\vec{\mathcal{O}}_3$ is determined as follows(the coordinate axis
$z$ is parallel to $\vec{\mathcal{O}}_3$):
\begin{eqnarray}
  \Phi_{mn}& = & \frac{\partial}{\partial \varepsilon_{zz}} \langle \psi_m | \hat{\mathcal{H}} | \psi_n \rangle = \nonumber \\
           & = & \frac{G_{33}}{2} \bigl( 3 \langle \psi_m | \hat{S}_z^2 | \psi_n \rangle - S(S+1) \langle \psi_m | \psi_n \rangle \bigr).
  \label{eq:05}
\end{eqnarray}
Here, $\varepsilon_{zz}$ is the component of the elastic strain
tensor; $\hat{\mathcal{H}}$ is the Hamiltonian of spin-phonon
interaction \cite{b07-TuckerE-1966,b38-FTT-1973,b39-TuckerJ-1972};
$| \psi_m \rangle$ and $| \psi_n \rangle$ are the wave functions of
paramagnetic ion in the crystalline field (these wave functions
belong to the ion's energy levels ${\mathcal{E}}_m$ and
${\mathcal{E}}_n$) respectively; ${G_{33}}$ is the component of the
spin-phonon interaction tensor
\cite{b07-TuckerE-1966,b38-FTT-1973,b39-TuckerJ-1972}; and
$\hat{S}_z$ is the projection of the vectorial spin operator onto
the $z$-axis.

In order to estimate $\Phi_{mn}$, let us use the experimentally
found value $G_{33} = 5.8$~cm${}^{-1}$ = $1.16 \cdot 10^{-15}$~erg
\cite{b38-FTT-1973}, as well as the wave functions $| \psi_3
\rangle$ and $| \psi_ 2 \rangle$ which belong to the energy levels
${\mathcal{E}}_3$ and ${\mathcal{E}}_2$ of the ${\mathrm{Cr}}^{3+}$
ion in the trigonal crystalline field of ruby (in this work,
similarly to works \cite{b21-TP-2004,b22-cond-mat-0402640}, $m =3;
n = 2$ ). From Eq.~(\ref{eq:05}), for the magnetic field $H =
3,92$~kOe directed at an angle $\vartheta =
\vartheta_{\mathrm{symm}}$ with respect to the $z$-axis, where
$\vartheta_{\mathrm{symm}} \equiv \arccos(1/\sqrt{3}) =
54^\circ44'$, we find $\Phi_{32} \approx 10^{-15}$~erg. The choice
of $\vartheta = \vartheta_{\mathrm{symm}}$ was dictated by the
requirements of the so-called symmetric (or push-pull) pumping
regime which was also used by us earlier to enhance the inversion
ratio in a phaser \cite{b21-TP-2004,b22-cond-mat-0402640}. As the
result, at $\nu = 9{.}1$~GHz, $g(\nu) = 10^{-8}$~s, $C_a = 1.3
\cdot 10^{19}$~cm${}^{-3}$, $\rho^{\prime} = 4$~g/cm${}^3$, $V_L =
1{,}1 \cdot 10^6$~cm/s, $T = 1{.}8~{\mathrm{K}}$, we find from
Eq.~(\ref{eq:04}) that $\sigma_{mn} = \sigma_{32} \approx
0{.}04$~cm${}^{-1}$.

The acoustic Q-factor $Q_C^{(1)}$ for our ruby AFPR (loaded by the
piezoelectric film) was measured making use of the pulse-echo
method at the frequency $\omega = 9{.}12$~GHz. The value of
$Q_C^{(1)}$ was found to be $5{.}2 \pm 0{.}4) \cdot 10^5$ at
$\vec{H} = 0$ and $P = 0$. Whence, $\eta \equiv
\eta_{\mathrm{vol}}$ + $\eta_{\mathrm{mirr}} = \omega /
Q_C^{(1)}V_L  \approx 0{,}1~{\mathrm{cm}}^{-1}$.

For the case of an autonomous phaser, a simple relation takes place
between $\sigma_{32}$, $\eta$ and the threshold value
$\alpha_g^{(1)}$, at which the generation of the first mode begins.
This relation reads as follows:
\begin{equation}
  \alpha_g^{(1)} = \eta = K_g\sigma_{32},
  \label{eq:06}
\end{equation}
where $K_g$ is the critical value of the inversion ratio $K$ at the
transition ${\mathcal{E}}_3 \leftrightarrow {\mathcal{E}}_2$.
Substituting $\sigma_{32} \approx 0.04$~cm${}^{-1}$ and $\eta
\approx 0.1$~cm${}^{-1}$ into Eq.~(\ref{eq:06}), we find that $K_g
\approx 2.5$, which can be ensured liberally by the push-pull
pumping scheme.

The autogeneration of longitudinal hypersound in the autonomous
ruby phaser was detected provided that external perturbations were
absent both in the signal and pump channels. The microwave
electromagnetic signal, excited in the hypersound transducer by the
microwave phonon SE, was supplied to a heterodyne spectrum
analyzer, and the spectra obtained were registered from its screen
by a photocamera. All these experiments were fulfilled at
temperatures below the superfluid critical point of liquid helium
$\theta < 2{.}1$~K, which allowed us to avoid the difficulties
related to the boiling of this cryogenic liquid.

Now, let us consider the obserwed microwave power spectra of the
phonon SE in the autonomous phaser generator in detail.

\section{Coarse Structure of the Spectra of Stationary
Phaser Autogeneration}\label{se:3}

Since the frequency width $\Gamma_{s}$ of the APR line for the spin
transition $E_3 \leftrightarrow E_2$ amounts to approximately
$100$~MHz, and the distance between the AFPR modes is only about
$300$~kHz, the primary single-mode SE easily transfroms to a
multi-mode one, even if the excess over the pumping threshold is
comparatively small. In other words, there emerges a coarse
structure (CS) of the phonon generation (see Fig.~1). Provided
$\omega_{\mathrm{pump}} = \omega_{cp}^{(0)} = 23.0$~GHz, and $H =
H_0 = 3.92$~kOe, the multimode phonon generation is observed even
at $P \ge 50 \, {\mu} {\mathrm{W}}$.

\vspace{10pt}

\begin{center}
  \framebox{\makebox{Figure 1 near here}}
\end{center}

\begin{quote}
\small{\textbf{Fig.~1:} Microwave power spectrum of longitudinal
phonon SE at $\Delta_P = \Delta_H = 0$ in the ruby phaser. The
frequency interval between neighboring CS modes of phonon SE
amounts to 310~kHz. The cryostat temperature is $\theta = 1.7$~K.}
\end{quote}


\vspace{10pt}

Under the condition of the magnetic-field detuning $\Delta_H \equiv
H - H_0 \neq 0$, it is natural that a considerably more intense
pumping is needed for the condition $K > K_g$ to be satisfied. On
the other hand, the increasing of a pumping level is also necessary
if there is some frequency detuning $\Delta_P \neq 0$, where
$\Delta_P \equiv \omega_{\mathrm{pump}} - \omega_{cp}^{(0)}$.

If the pumping is swithed on abruptly, the transient process, which
takes place in the course of establishment of the stationary
integral intensity of SE $J_\Sigma$, demonstrates an oscillatory
behavior. The frequency $\omega_{\nu}$ of these damped oscillations
(the so called relaxation frequency) for our system has, as it was
already mentioned, a value of the order $\omega_{\nu} / 2 \pi
\approx 10^2$~Hz \cite{b23-TP-1989,b24-TP-1992}.

The lifetime of this transient process $\tau_{\mathrm{tran}}$ is
shorter than $0{,}3$~s at $P > 0{,}1$~mW and provided that the
magnetic-field and pump-frequency detunings are absent ($\Delta_H =
0, \Delta_P =0$). Therefore, if the phaser system is finely tuned
and the pumping is powerful enough, the transient process is rather
fast ($\tau_{\mathrm{tran}} \approx \tau_1$). This experimental
fact agrees with theoretical estimations
\cite{b23-TP-1989,b24-TP-1992} of $\tau_{\mathrm{tran}}$ made in
the framework of the elementary (balance) model of SE
\cite{b41-TCRE-2002,b42-nlin-0704.0123}: $\tau_{\mathrm{tran}}
\approx \tau_1 / A_P$, where $A_P$ is the pumping parameter (for
our system, $A_P \approx 2$ \cite{b23-TP-1989,b24-TP-1992}). In
other words, under the conditions specified above, the process of
formation of the stationary CS of the phaser power spectra
terminates approximately within the same time interval, over which
the stationary resonant absorption in a passive (non-inverted)
paramagnetic system becomes settled.

\section{Regular Fine Structure and Slow Transient Processes
in the Microwave Power Spectra}\label{se:4}

\subsection{Conditions of emergence of a regular fine
structure}\label{subse:4-1}

At $P \geq 4$~mW, $\Delta_P =0$, and the small magnetic-field
detunings $|\Delta_H| \leq 2 - 3$~Oe, the integral intensity
$J_\Sigma$ of multimode SE of an autonomous phaser, as it was
already pointed out in works
\cite{b21-TP-2004,b22-cond-mat-0402640}, practically does not
depend on time. Provided that the pumping is powerful enough and
its frequency is tuned finely, the measured magnitude of $J_\Sigma$
weakly (in the limits of several percent) oscillates only if
$|\Delta_H|$ is increased to about $\approx 30$~Oe and more
\cite{b21-TP-2004,b22-cond-mat-0402640}.

In the corresponding non-stationary microwave phonon power spectra,
some CS modes of autonomous phaser become splitted at $|\Delta_H|
\geq 30$~Oe. One of the typical cases of such the splitting for a
phaser SE mode is shown in work \cite{b43-UJP-2002}. Thus, the fine
structure (FS) spontaneously arises in microwave phonon power
spectrum. We stress that no external forcing is needed for
emergence of FS, in contrary to other nonlinear phenomena
investigated by us at the CS level
\cite{b14-TPL-2001,b15-cond-mat-0303188,b21-TP-2004,b22-cond-mat-0402640,b43-UJP-2002}.
The intensities of the FS components $J_F^{(i)}$ observed in
\cite{b43-UJP-2002}, were very low, even at $i_{\max} = 2-3$. This
is several orders of magnitude lower than the intensities of the
non-splitted SE components of stationary generation which are
depicted at Fig.~1.

However, if the phaser system becomes appreciably detuned by the
pump frequency ($\Delta_P >> \Delta_n \approx 0{,}3$~MHz), the
emergence of the FS in the microwave power spectra of phaser
generation is observed even at $\Delta_H = 1-2$~Oe. In this case,
$J_F^{(i)}$ are of the same order of magnitude as the unsplitted CS
components. In Fig.~2, the splitting of the CS mode into two and
three intense FS components is demonstrated, as $\Delta_H$
increases from 0 up to $3{,}5$~Oe. One can also notice very weak
additional FS components against the noise background.

\vspace{10pt}

\begin{center}
  \framebox{\makebox{Figure 2 near here}}
\end{center}

\begin{quote}
\small{\textbf{Fig.~2:} Emergence of a regular FS in the power
spectra of an autonomous phaser at $P = 4$~mW, $\Delta_P =
8.8$~MHz, $\theta = 1.8$~K, and for various $\Delta_H$. Left
oscillogram: $\Delta_H = 0$ (the FS is absent). Middle oscillogram:
$\Delta_H = $~1,5~Oe (the two-component FS). Right oscillogram:
$\Delta_H = $~3,5~Oe (the three-component FS). The range of the
frequency sweeping along the horisontal axis is about 6~kHz for
each oscillogram.}
\end{quote}


\vspace{10pt}

\subsection{Mechanism of formation of a regular
fine structure}\label{subse:4-2}

The appearance of the regular FS with a few (2 to 3) components of
the splitted CS mode of phonon SE can be explained in the framework
of the Casperson--Yariv mechanism \cite{b44-Harrison-1985}. The
normalized dispersion $\Delta V_L / V_L$ of the phase velocity for
hypersound with the frequency $\omega$ in a paramagnetic AFPR
possessing a certain eigenfrequency of the CS mode $\omega_{cs}
\approx \omega $ looks like
\begin{equation}
  \frac{\Delta V_L}{V_L} = \frac{\alpha \xi_r }{ 2k_L (1 + \xi_r^2) }
  = \frac{q_M \xi_r }{ (1 + \xi_r^2)} ,
  \label{eq:07}
\end{equation}
where $\xi_r \equiv (\gamma_{s} H_0 - \omega) / \Gamma_{s}$,
$\gamma_{s} = {\hbar}^{-1} \left [
\partial ({\mathcal{E}}_3 - {\mathcal{E}}_2) / \partial H \right
]$, $H_0 = \omega_{cs} / \gamma_{s} = \pi {n} V_L / \gamma_{s} L $,
and $q_M = \alpha / 2k_L$.

Now, let us introduce the additional detuning of the system by
changing the static magnetic field $H$ with respect to its resonsnt
value $H_0$. Using the dimensionless magnetic-field detuning $h =
\gamma_{s} \Delta_H \Gamma_{s}^{-1} = (\gamma_{s} H - \omega_{cs})
/ \Gamma_{s}$, we have the following formula for the dependence of
the dispersion of the hypersound phase velocity on the magnetic
field:
\begin{equation}
  \frac{\Delta V_L (h)}{V_L} = \frac{ q_M \delta }{ (1 + \delta^2) },
  \label{eq:08}
\end{equation}
where
\begin{equation}
  \delta \equiv \xi_r + h = \frac{\gamma_{s} H -
  \omega}{\Gamma_{s}}.
  \label{eq:09}
\end{equation}

Now, we apply the well-known Casperson--Yariv relation
\cite{b44-Harrison-1985} which couples the refraction index $\eta$
(in our case, this is the acoustic refraction index $\eta =
\eta_{\mathrm{acoust}}$) with the mode frequency
$\overline{\omega}$ in the resonator (in our AFPR, this is a
hypersonic mode, for which Eqs. (\ref{eq:08})--(\ref{eq:09}) hold
true):
\begin{equation}
  (\eta_{\mathrm{acoust}} - 1) L_A \overline{\omega} / L =
  \omega_{cs}- \overline{\omega},
  \label{eq:10}
\end{equation}
where $L_A$ is the active medium length. In the used phaser system,
$L_A = L$; therefore, making use of the relation $\Delta V_L / V_L
= 1 - \eta_{\mathrm{acoust}}$ and formulae (\ref{eq:08}) and
(\ref{eq:09}), we obtain the implicit expression for calculating
the dependence $\delta(h)$:
\begin{equation}
  h - \delta(h) = \frac{q_M \overline{\omega}(h)}{\Gamma_{s}} \cdot
  \frac{\delta(h)}{1+\delta^2(h)}.
  \label{eq:11}
\end{equation}

On the right-hand side of Eq. (\ref{eq:11}), we may put
$\overline{\omega} = \omega_{cs}$. Then, from Eq. (\ref{eq:11} and
using the equality $\xi_r = \delta - h$, we find the FS spectrum
for the AFPR mode as a function of the magnetic field:
\begin{equation}
  \overline{\omega}^{(j)}(h) = \omega_{cs} + \left [ h -
  \delta^{(j)} (h) \right ] \Gamma_{s}.
  \label{eq:12}
\end{equation}
Here, $\delta^{(j)}$ are the roots of the Bonifacio--Lugiato
equation \cite{b45-Gibbs-1985}:
\begin{equation}
  h = \left ( 1 + \frac{2M}{1 + \delta^2} \right ) \delta,
  \label{eq:13}
\end{equation}
and the control parameter $M$ is of the form
\begin{equation}
  M = \frac{q_M \omega_{cs}}{2 \Gamma_{s}} \propto
  \frac{K C_a k_u^2 |\Phi_{32}|^2}{\Gamma_s^2 k_B \theta}.
  \label{eq:14}
\end{equation}
At high magnetic-field detuning $h$, we have $\overline{\omega}
\rightarrow \omega_{cs}$; while at $h \rightarrow 0$ we see that $|
\overline{\omega} - \omega_{cs} | \propto h$. Accordingly, the
function $\delta(h)$ is single-valued for these extrme cases. In
the intermediate range of magnetic-field detunings $h$, the
function $\delta(h)$ may be multi-valued. Bifurcation points which
determine conditions of this branching of $\delta(h)$ (i.~e.
splitting of a CS mode into two or three FS components) can be
easily calculated by standard methods of the catastrophe theory
\cite{b46-Poston-1978}. We find two bifurcations of codimension 2
(at $M = M^{(0)} = 4$ and $h = h_{\pm}^{(0)} = \pm 3 \sqrt{3}$) and
four bifurcations of codimension 1, which take place at $M
> M^{(0)}$, $h = h_{\pm}^{(\pm)} = \pm \left \{(M/2) \left [ a(M)
\pm b(M) \right ] \right \}^{1/2}$. Here signs in subscript
correspond to signs at curly brackets, signs in superscript
correspond to signs within square brackets, $a(M) = a + 10 -
2a^{-1}$ and $b(M) = [(a-4)^3 /a]^{1/2}$. So, the function
$\delta(h)$ is two-valued at $M = M^{(0)}$ or at $M
> M^{(0)}$ and $h = h_{\pm}^{(\pm)}$. Accordingly, the function
$\delta(h)$ is three-valued if $M > M^{(0)}$ and $h_{+}^{(-)} < h <
h_{+}^{(+)}$ or $h_{-}^{(+)} < h < h_{-}^{(-)}$. This
three-component splitting is typical due to it's non-critical
nature. The two-component splitting is, strictly speaking,
critically dependent on variations of $h$, but in real experiment
there exist a small but finite range of $h$ in the vicinity of
bifurcation point making it possible to observe unhesitatingly the
two-component FS (see Fig.~2, middle oscillogram).

Thereby, the ``static'' Casperson-Yariv model considered above
demonstrates a possible mechanism of spontaneous (not connected
with external factors) splitting of a CS mode into two or three
components, depending on magnetic field detuning. This model gives
a satisfactory interpretation of such the splitting FS. Moreover,
the Casperson-Yariv model can be improved considerably by making
allowance for the effects of hole burning in the amplification line
etc. \cite{b44-Harrison-1985}. But a series of other results on
microwave phonon SE in the autonomous phaser needs alternative
approaches for at least qualitative explanation of phaser dynamics
observed in our experiments. Very interesting, in particular, are
the issues concerning transient processes of the FS emergence,
especially in the case where the characteristic time of a transient
process $\tau_{\mathrm{tran}}$ is much longer than the spin-lattice
relaxation time $\tau_1$ (and a control parameter set is far from
the critical one). Processes of self-induced non-critical slowing
of evolution have been studied intensively in recent years in
nonlinear autonomous systems of various nature, and important
results have already been obtained by means of computer modeling
\cite{b30-Morita-2004,b31-Morita-cond-mat-0304649,b32-Morita-cond-mat-0407460,b33-Awazu-2004,b34-Awazu-nlin-0310018,b35-SDM-cond-mat-0602345,b36-SDM-Master-Thesis-2006,b37-SDM-cond-mat-0410460}.
In this work, we focus our attention first on two specific issues,
namely, whether such slow processes may proceed in an autonomous
phaser and which is the nature of coexistence of incongruous states
arising during the pointed transient processes in the phaser active
medium. The relation between obtained experimental results and
those of computer modeling
\cite{b30-Morita-2004,b31-Morita-cond-mat-0304649,b32-Morita-cond-mat-0407460,b33-Awazu-2004,b34-Awazu-nlin-0310018,b35-SDM-cond-mat-0602345,b36-SDM-Master-Thesis-2006,b37-SDM-cond-mat-0410460}
will be discussed in details after exposition of experimental data.

\subsection{Slow transient processes in phaser during formation
of the regular fine structure}\label{subse:4-3}

Our observations of the spectra of phonon SE in the autonomous phaser showed
that, provided $\Delta_H \neq 0$ and $\Delta_P \neq 0$, the characteristic time
of transient processes, which accompany the emergence of the spectral FS,
considerably exceeds the time of longitudinal relaxation of
$\mathrm{{Cr}^{3+}}$ ions in ruby at the same temperature $\theta$:
$\tau_{\mathrm{tran}}(\theta)
>> \tau_1(\theta)$. For example, for the FS depicted in Fig.~2,
$\tau_{\mathrm{tran}}(\theta_0) \approx 30$~s (here, $\theta_0 = 1{,}8$~K). Not
only is this by two orders of magnitude longer than $\tau_1(\theta_0)$, but
also is much longer than the time of establishing of the stationary CS at
$\Delta_H = 0$ and $\Delta_P =0$. Moreover, after the process having
terminated, the structure of the power spectra is, as a rule, nonstationary.
Instead, there appear periodic oscillations of the intensities of the FS
components or even the periodic motions of these components along the frequency
axis whithin the limits of several kHz. These oscillations and motions, as well
as the preceding transient process, run slowly in comparison to the relaxation
time of active $\mathrm{Cr}^{3+}$ centers.

At first glance, the anomalously great values of
$\tau_{\mathrm{tran}}$ which were observed in the generating phaser
system under conditions of detunings $\Delta_H \neq 0, \Delta_P
\neq 0$ can be explained as the result of influence of the slowly
relaxing subsystem of magnetic nuclei ${}^{27}\!\mathrm{Al}$ of
aluminium ions $\mathrm{Al}^{3+}$ which belong to the corundum
crystalline matrix. Such an influence of ${}^{27}\!\mathrm{Al}$
nuclei on electron paramagnetic system of active $\mathrm{Cr}^{3+}$
centers has been revealed experimentally earlier
\cite{b12-JETP-1977,b28-Diss-1983,b29-Aref-1984} at qualitatively
different conditions, namely in experiments on phaser
{\textit{amplification}} of hypersound in ruby. Studying of phaser
amplification in \cite{b12-JETP-1977,b28-Diss-1983,b29-Aref-1984}
were fulfilled, naturally, below the self-excitation threshold of
phaser generation, i.~e. at inversion ratios $K < K_g$. In this
case, the dominant role of electron-nuclear interactions in
slowing-down of transient processes of non-generating phaser was
established in \cite{b12-JETP-1977,b28-Diss-1983,b29-Aref-1984} and
confirmed in alternative experiments on phaser amplification at the
mentioned condition $K < K_g$ (under condition of self-focusing of
hypersound) \cite{b47-FTT-1986}. However, our further researches
have demonstrated that under the phaser {\textit{generation}}, when
$K > K_g$, this electron-nuclear mechanism of transient processes
slowing-down does not dominate.

Really, the slow motions, which were observed in the
electron-nuclear magnetic system of ruby phaser amplifier
\cite{b12-JETP-1977,b28-Diss-1983,b29-Aref-1984}, are characterized
by the well-known restriction on the maximal time of the transient
process $\tau_{\mathrm{tran}}$ in this system --- the relaxation
time $\tau_{\mathrm{nucl}}^{(z)}$ of the Zeeman reservoir for the
subsystem of magnetic nuclei
\cite{b48-Atsarkin-1978,b49-Abragam-1982}. Under our experimental
conditions, $\tau_{\mathrm{nucl}}^{(z)}$ does not exceed 10~s,
which is three times shorter than the values of
$\tau_{\mathrm{tran}}$ observed at the FS emergence in the power
spectra of the phaser generation.

More detailed researches of the power spectra of an autonomous phaser
demonstrated that, under definite conditions, the FS formation can occur
following much more comlcated scenarios than those in the experiments described
above. In this case, the ultimate state of the SE line turns out qualitatively
different from what is shown in Fig.~2, and the time of the corresponding
transient process grows substantially ($\tau_{\mathrm{tran}} > 10^2$~s). Such
the situation takes place, e.~g., if, provided $\Delta_P >> \Delta_n$, the
parameter $\Delta_H$ is increased several times with respect to those
$\Delta_H$ values, for whch the FS shown in Fig.~2 was observed. Let us
consider these experimens in detail.

\section{Super-slow Transient Processes and Selective Chaotization
of Fine Structure in Microwave Phonon Power Spectra}\label{se:5}

\subsection{Experimental observation of selective chaotization
of fine structure}\label{subse:5-1}

Provided the same values of pump power and pump frequency detuning
as in the previous experiments (i.~e. $P = 4$~mW, $\Delta_P
=$~8,8~MHz), but for the enlarged magnetic-field detuning up to
$\Delta_H = $~15~Oe, the regular FS (which is shown in  Fig.~2) was
found to be gradually destroyed. In approximately 10~min after the
indicated detuning having been introduced, the form of some FS
lines became apparently chaotic. In this work, we use the term
``chaotic FS'', not specifying the kind of the disorder observed,
because it is not still clear, whether this is a multi-dimensional
deterministic chaos or some form of a small-scale spin-phonon
turbulence. The experimental data presented below allow us to
assert only that the dimension of the phase space, which is
necessary to describe the observed chaotic FS, should be much
higher than that in the case of the ordinarylow-dimension
deterministic chaos which was studied in phasers whith periodic
pump modulation
\cite{b23-TP-1989,b24-TP-1992,b41-TCRE-2002,b42-nlin-0704.0123}.

A typical view of the chaotic FS in an autonomous phaser is shown
in Fig.~3. Unlike the regular FS which can be either static or
periodically pulsing, the chaotic FS is distinguished for fast,
irregular, and mutually non-synchronized pulsations of the
amplitudes of its numerous components. Moreover, these chaotic
pulsations of amplitudes are accompanied by similar irregular and
mutually non-synchronized motions of the FS components along the
frequency axis (back and forth in narrow frequency windows).

\vspace{10pt}

\begin{center}
  \framebox{\makebox{Figure 3 near here}}
\end{center}

\begin{quote}
\small{\textbf{Fig.~3:} Chaotic FS of the phonon generation at
$\theta = 1.8$~K, $P = 4$~mW, $\Delta_P = 8.8$~MHz, and $\Delta_H =
15$~Oe. The range of the frequency sweeping along the horisontal
axis is about 18~kHz.}
\end{quote}


\vspace{10pt}

From Fig.~3, one can see that the spectral width of such a chaotic
FS line of the phonon SE is several times greater than the width of
the line with a regular FS. At the same temperature $\theta =
1.8$~K, holding both $\Delta_P$ and $\Delta_H$ constant (namely,
$\Delta_P =$~8.8~MHz, $\Delta_H = $~15~Oe), but reducing the pump
power $P$ by a factor of four, a diminishing of the peak intensity
of the chaotic FS components was observed, as well as some
narrowing of the FS line as a whole (Fig.~4).

\vspace{10pt}

\begin{center}
  \framebox{\makebox{Figure 4 near here}}
\end{center}

\begin{quote}
\small{\textbf{Fig.~4:} The same as in Fig.~3, but at reduced pump
power $P = 0.9$~mW.}
\end{quote}


\vspace{10pt}

The further reduction of $P$ results in disruption of the phonon
generation at the chaotic SE mode. Just before this mode
extinguishing (at $P \approx 0.1$~mW), its structure becomes
somewhat ordered as compared with cases shown at Fig.~3 and Fig.~4.

\subsection{Superslow transient processes at fine structure
chaotization}\label{subse:5-2}

The disruption of the phonon generation at a chaotic SE mode may
happen, as it was observed in our experiments, at fixed $P$ as
well, but when varying $\Delta_P$ or $\Delta_H$. In so doing,
contrary to the case of reducing $P$, the chaotization of the
nearest neighboring unsplitted CS mode was often observed.

It is essential that the described phenomena of chaotization of FS
lines are very selective with regard to detunings of the phaser on
the pump frequency and the static magnetic field. Chaotization take
place only in the rather narrow windows of $\Delta_P$ and
$\Delta_H$. Outside of these windows, the strongly marked
chaotization of the SE lines was not achieved, but some
intermediate states of the FS were observed outside such windows as
well. For example, a FS line may contain only two or three robust
components, which pulse smoothly and move asynchronously along the
frequency axis (within an interval of about 20~kHz). A kind of such
an intermediate FS state (emerged in the same ruby phaser, but at
$\Delta_H \approx $~30~Oe) was already exemplified by us in
\cite{b21-TP-2004,b22-cond-mat-0402640}.

The described above internally caused chaotic pulsations of the FS
components in our autonomous phaser are fast for most fixed sets of
the control parameters used in our experiments. Typical times of
the FS intramode motions are less than 1~s. On the other hand, the
externally excited \textit{transitions} from one chaotic state to
another (initiated by a step-type changing of at least one of the
control parameters) proceed even more slowly than in the case of
transitions between various states of a regular FS. Typical values
of $\tau_{\mathrm{tran}}$ for transitions between states of a
chaotic FS (e.g., between states shown at Fig.~3 and Fig.~4) amount
to $\tau_{\mathrm{tran}} \gtrsim 10^3$~s. No external forcing of
the system takes place here, so we obviously deal with
self-organized bottleneck in transients between selective chaotic
states in phaser medium.

Giant times of transient processes are comparable with the
characteristic times of superslow motions observed earlier in the
non-autonomous (periodically-modulated) ruby phaser
\cite{b21-TP-2004,b22-cond-mat-0402640}. However, the similarity
between superslow processes in non-autonomous (studied in
\cite{b21-TP-2004,b22-cond-mat-0402640}) and autonomous (studied in
the present paper) phasers at this point comes to end. For the
\textit{non-autonomous} phaser
\cite{b21-TP-2004,b22-cond-mat-0402640}, the emergence of superslow
motions is a characteristic feature of the whole power spectrum
(the typical width of the phonon SE power spectrum amounts to
several MHz). In this case, the ``firing'' of microwave phonon CS
modes on one side of the spectrum is accompanied by the
``extinguishing'' of approximately the same number of CS modes on
its opposite side. This phenomenon of alternation of microwave
phonon CE modes was observed in
\cite{b21-TP-2004,b22-cond-mat-0402640} under deep modulation of
phaser pump at frequencies of nonlinear resonance (about $9.8$~Hz
in the investigated system) and its even harmonics.

Of the \textit{autonomous} phaser, typical are the scenarios, when
the majority of microwave phonon CS modes are almost stationary,
and their frequencies are very close to frequencies of normal modes
of the AFPR. Only the small quantity of the CS modes (usually no
more than two), as it was described above, are splitted, shifted in
frequency and even chaotized at certain combinations of the control
parameters. The definite modification of such a combination of the
control parameters is needed in order to translocate this selective
phonon mode ``fever'' to the nearest neighboring CS modes. But
times of the mode structure reconfiguration are much more greater
than times of the individual AC's relaxaton time. Generally
speaking, there are two unusual and interrelated phenomena observed
in our autonomous system: coexistence of incongruous states
(neighboring of dominant normal and selectively chaotized CS modes)
and the self-organized bottleneck in transients between these
states in phaser medium.

\section{A Possible Direction for Numerical Modeling of Self-Organized
Bottleneck and Coexistence of Incongruous States in Phaser and
Other Excitable Systems}\label{se:6}

\subsection{Macro- and microscopic approaches to simulating
nonlinear phenomena in systems of the phaser type}\label{subse:6-1}

A theoretical model of the observed highly nonlinear phenomena
requires a separate studies which must include spatio-temporal
microscopic description of the local interactions between
individual ACs. These interactions were considered by us earlier
\cite{b12-JETP-1977} only at the level of nonequilibrium
thermodynamics, namely, by using spin-temperature approach
\cite{b48-Atsarkin-1978,b49-Abragam-1982}, which by definition
neglects individuality of the ACs.

Some preliminary numerical results towards this direction have been
obtained in
\cite{b35-SDM-cond-mat-0602345,b36-SDM-Master-Thesis-2006,b37-SDM-cond-mat-0410460,b50-SDM-nlin-0805.1319}.
Below, the spatial phenomena and the transient processes, which
were observed in computer experiments on excitable systems
\cite{b35-SDM-cond-mat-0602345,b36-SDM-Master-Thesis-2006,b37-SDM-cond-mat-0410460,b50-SDM-nlin-0805.1319},
as well as their relation to the described above results of real
experiments (including self-organized bottleneck and coexistence of
incongruous states in the autonomous phaser), will be discussed.
But first, it is necessary to stop at some conceptual moments
dealing with phaser models.

Slow transient processes at the phaser \textit{amplification} of
hypersound (below the threshold of the phaser self-excitation) were
observed in 1970--1980
\cite{b12-JETP-1977,b28-Diss-1983,b29-Aref-1984}. The
characteristic time of those processes did not exceed, as it was
said above, the relaxation time of ${}^{27}\!\mathrm{Al}$ nuclei
which form an additional energy reservoir in the phaser active
medium. This reservoir interacts with the electron spin subsystem
of the active cromium centers in the crystalline matrix of
$\mathrm{Al}_2\mathrm{O}_3$ owing to a thermal contact between the
system of ${}^{27}\!\mathrm{Al}$ nuclei and the electron
dipole-dipole ($d$--$d$) reservoir made up by locally interacting
$\mathrm{Cr}^{3+}$ \cite{b48-Atsarkin-1978,b49-Abragam-1982}.

Such a contact is possible due to the proximity of the nuclear
magnetic resonance (NMR) frequencies of ${}^{27}\!\mathrm{Al}$
\textit{nuclei} to the characteristic frequencies of
\textit{electronic} $d$--$d$ interactions in the system of
$\mathrm{Cr}^{3+}$ ions in ruby (all these frequencies are of the
order of 10~MHz). As a result, the slow evolution of the joint
electron-nuclear system  of thew ruby phaser amplifier can be
described adequately
\cite{b12-JETP-1977,b28-Diss-1983,b29-Aref-1984} in the framework
of the nonequilibrium spin thermodynamics concept
\cite{b48-Atsarkin-1978,b49-Abragam-1982}.

It is known that those concepts are based on the hypothesis about
the so-called spin temperatures
\cite{b48-Atsarkin-1978,b49-Abragam-1982}, i.~e. some macroscopic
characteristics of the energy reservoirs of a nonequilibrium
quantum system, the saturation of which is typically ensured by an
external coherent field with a frequency close to one of the
system's resonances (NMR, ESR, or APR). We emphasize that it is the
fact of saturating the system by an external resonant field that
enables us to analyze the distribution of spin level populations in
the coordinate system which rotates synchronously with the
frequency of the applied field. Under rather general conditions,
such a distribution of spin-level populations is exponential, which
allows us to speak about such a macroscopic parameter as the spin
temperature.

A different is the situation, when various modes of phaser
\textit{generation} emerge, evolve, and reach their dynamically
stable states. Really, even provided that the FS is absent, the
spin system of a phaser generator becomes saturated not only due to
the pump field, but also due to the fields of microwave phonon SE
at several CS frequencies (see Fig.~1), because the single-mode
phaser generation is unstable
\cite{b09-SSC-1974,b12-JETP-1977,b28-Diss-1983,b29-Aref-1984}. In
this case, individual spin temperatures for all the individual spin
reservoirs (which generate the corresponding CS microwave modes)
have to be considered. The quantity of simultaneously
self-saturating microwave CS modes with various eigenfrequencies
reaches about 10 -- 20 and more
\cite{b09-SSC-1974,b12-JETP-1977,b28-Diss-1983,b29-Aref-1984}. So
the thermodynamical approach is here not so clear as in the case of
single-frequency saturation studied in
\cite{b48-Atsarkin-1978,b49-Abragam-1982}.

Again, the occurence of a regular FS (Fig.~2) not only increases
the number of macroscopic parameters, which are necessary for
description of the active system behavior, but also calls into
question the adequacy of such a thermodynamical approach in
general. Concerning the \textit{chaotic} FS of a phaser generator
(Fig.~3), the macroscopic thermodynamical model of saturated spin
system \cite{b48-Atsarkin-1978,b49-Abragam-1982}, which yelded the
satisfactory results for a phaser amplifier
\cite{b12-JETP-1977,b28-Diss-1983,b29-Aref-1984}, becomes obviously
unacceptable now.

In order to understand the mechanisms of formation of the chaotic
FS (including such issues as the slowing-down of transient
processes, the coexistence of regular and chaotic CS modes, and so
on) we have to deal with the simulation of inversion states
dynamics at the \textit{microscopic} level.

Carrying out the direct simulation of the evolution of an active
multiparticle system on the basis of the Maxwell--Bloch coupled
equation \cite{b51-Ikeda-1989} becomes inexpedient even at the
number of ACs $N \gtrsim 10^3$, because too huge computational
resources would be required for this purpose. Alternative is the
way to apply discrete models of cellular automata (CA) type
\cite{b52-Wolfram-2002}. Such models use algorithmic schemes with
local information processing ($K$- or Kolmogorov algorithms
\cite{b53-Kolmogorov-1963} and allow the corresponding CA-programs
to run effectively on ordinary personal computers even if $N
\gtrsim 10^6$. It is important that $K$-algorithms used in most
CA-models belong to the $P$-time class of complexity
\cite{b54-Cormen-2001}, which allows $N$ to be increased by several
orders for middle-class computers.

\subsection{Possibility of the formation of a self-organized bottleneck
in active systems of the phaser type}\label{subse:6-2}

In works
\cite{b35-SDM-cond-mat-0602345,b36-SDM-Master-Thesis-2006,b37-SDM-cond-mat-0410460,b50-SDM-nlin-0805.1319},
a series of computer experiments on a three-level CA (TLCA) model
of a class-$B$ phaser-like system was carried out. The prototype
model (Zykov-Mikhailov model) \cite{b55-Zykov-1986} of excitable
system includes mechanisms of global activation (GA), global
inhibition (GI) and local activation (LA) of ACs. The LA mechanism
is a deterministic analog of the one-channel (1C) diffusion of
excitations. The TLCA model proposed in
\cite{b35-SDM-cond-mat-0602345,b36-SDM-Master-Thesis-2006,b37-SDM-cond-mat-0410460,b50-SDM-nlin-0805.1319}
takes into consideration local inhibition (LI) too. In the
framework of the TLCA model
\cite{b35-SDM-cond-mat-0602345,b36-SDM-Master-Thesis-2006,b37-SDM-cond-mat-0410460,b50-SDM-nlin-0805.1319},
the combined LA/LI interaction between ACs is a deterministic
analog of the two-channel (2C) diffusion of excitations. It is well
known that in dilute paramagnets the main mechanism of local
interactions between ACs is namely LA~\&~LI one. More concrete,
this is dipole-dipole magnetic interactions between impurity
paramagnetic ACs \cite{b26-Abragam-1970,b27-Altshuler-1972}.
Mechanisms of LA and LI are equal partners in dilute paramagnets,
in contrary to negligible role of LI in chemical reactions of
Belousov-Zhabotinskii type.

The results of computer experiments showed
\cite{b35-SDM-cond-mat-0602345,b36-SDM-Master-Thesis-2006,b37-SDM-cond-mat-0410460,b50-SDM-nlin-0805.1319},
that a typical scenario of the evolution of an autonomous class-$B$
system with weak diffusion of excitations is the self-organized
formation of vortex-like coherent dissipative structures of the
rotating spiral wave (RSW) type. In the case of 2C-diffusion,
transient times for RSW dissipative structures stabilization reach
giant values (of about $10^4 - 10^6$ times more than individual
AC's relaxation time). This  self-organized coherent bottleneck
differs qualitatively from the well-known incoherent phonon
bottleneck, which is absent in dilute paramagnets used as phaser
active media.

In contrast to the phonon bottleneck or any other incoherent
bottleneck, the coherent bottleneck observed in computer
experiments
\cite{b35-SDM-cond-mat-0602345,b36-SDM-Master-Thesis-2006,b37-SDM-cond-mat-0410460,b50-SDM-nlin-0805.1319}
is the result of very slow emergence of dissipative spatio-temporal
structures. In some aspects, slowing-down phenomena observed in
\cite{b35-SDM-cond-mat-0602345,b36-SDM-Master-Thesis-2006,b37-SDM-cond-mat-0410460,b50-SDM-nlin-0805.1319}
are similar to self-induced slowing-down of transient processes
discovered computationally in both a system of interacting
oscillators
\cite{b30-Morita-2004,b31-Morita-cond-mat-0304649,b32-Morita-cond-mat-0407460}
and a system of reaction-diffusion type
\cite{b33-Awazu-2004,b34-Awazu-nlin-0310018}. The last system is
very similar to the one modeled in
\cite{b35-SDM-cond-mat-0602345,b36-SDM-Master-Thesis-2006,b37-SDM-cond-mat-0410460}.

It is interesting, that under conditions of strong competition of
LA and LI mechanisms of AC-AC interaction (2C-diffusion) the
unexpected phenomenon of RSW ``reflections'' was discovered in
three-level CA model of excitable phaser-like system
\cite{b35-SDM-cond-mat-0602345,b36-SDM-Master-Thesis-2006}. In
essence, each ``reflection'' of RSW is, as it was observed in
\cite{b35-SDM-cond-mat-0602345,b36-SDM-Master-Thesis-2006}, the
result of nonlinear transformation of the RSW's core in surface
layer. Such the phenomenon of RSW \textit{regeneration} at the
boundaries of an active medium leads to further increasing of
transient time in TLCA. A close phenomenon of optical vortex
``self-repairing'' was observed earlier in real physical
experiments \cite{b56-Vasnetsov-2000}.

Moreover, computer experiments
\cite{b35-SDM-cond-mat-0602345,b36-SDM-Master-Thesis-2006}
demonstrated \textit{replications} of RSWs in surface layer of
three-level active system with 2C-diffusion of excitations. In this
case, each collision of RSW with boundary generate several new
RSWs, which move from boundary into inner area of the system.
Self-induced replications of RSWs may lead to emergence of
transient chaos, when attractor becomes effectively unattainable by
any digital computer with limited resources.

Hence, plausible looks the assumption that phenomenon of a
self-organized bottleneck is the cause of slowing-down of the
transient processes in a phaser. The additional evidence for this
assumption is the result of computer experiment
\cite{b36-SDM-Master-Thesis-2006,b37-SDM-cond-mat-0410460}
demonstrating coexistence of regular and irregular spatio-temporal
structures in TLCA (transient analog of chimera states
\cite{b57-Kuramoto-2002,b58-Abrams-2004,b59-Abrams-2006}). This
phenomenon correlates with the coexistence between the regular and
chaotic FS modes of phaser generation described above. Certainly,
the possibility for dissipative stuctures of the RSW type to exist
in a real phaser system still demands additional researches, also
spiral structures have already be observed in dissipative nonlinear
optical systems
\cite{b02-Weiss-1999,b03-Weiss-2007,b60-Akhmanov-1992,b61-Huneus-2004,b62-Weiss-2004}.

A particular emphasis should be made that, as it was already
pointed out in
\cite{b30-Morita-2004,b31-Morita-cond-mat-0304649,b32-Morita-cond-mat-0407460},
the self-organized bottleneck has the origin qualitatively
different from that of well-known phenomenon of self-organized
criticality (SOC) \cite{b63-Bak-1988,b64-Jensen-1998}. The SOC
phenomenon arises only in non-autonomous dissipative systems (as a
nontrivial response to an external destabilizing force), whereas
the self-organized bottleneck
\cite{b30-Morita-2004,b31-Morita-cond-mat-0304649,b32-Morita-cond-mat-0407460,b33-Awazu-2004,b34-Awazu-nlin-0310018,b35-SDM-cond-mat-0602345,b36-SDM-Master-Thesis-2006,b37-SDM-cond-mat-0410460,b50-SDM-nlin-0805.1319}
is a result of internal processes running in the
\textit{autonomous} dissipative systems.

In this aspect, the processes of self0organization observed
experimentally in non-autonomous (our previous works
\cite{b14-TPL-2001,b15-cond-mat-0303188,b21-TP-2004,b22-cond-mat-0402640})
and autonomous (this work) phasers may also be examined from
qualitatively different points of view. Really, slow self-organized
motions in the spin-phonon system of a non-autonomous phaser
\cite{b14-TPL-2001,b15-cond-mat-0303188,b21-TP-2004,b22-cond-mat-0402640}
arize just under the influence of an external resonant
destabilizing force and dissappear after swithching the latter off.
A different picture was observed in this work for autonomous
phasers, because here the slowing-down of the transient processes
is self-induced and does not require any external perturbations. On
the other hand, in the case of a non-autonomous system similar to
that described in works
\cite{b14-TPL-2001,b15-cond-mat-0303188,b21-TP-2004,b22-cond-mat-0402640},
the slow motions proceed as long as the corresponding destabilizing
force is active, while the transient process in an autonomous
system has, of course, a finite (though, may be, very big)
characteristic time. As we have shown in this work, the asymptotic
state of an autonomous phaser need not necessarily be regular,
because the FS may possess a very complicated, turbulent-like
structure.

\section{Conclusions}\label{se:7}

The structure and evolution of power spectra in a ruby-based
($\mathrm{{Cr}^{3+}} \: {:} \: \mathrm{{Al}_2{O}_3}$) autonomous
microwave phonon laser (phaser) have been studied experimentally at
liquid helium temperatures.

The self-organized bottleneck and coexistence of incongruous states
(regular and irregular phonon modes) were revealed during
investigation of FS of microwave phonon power spectra at
simultaneous detunings of both the pump frequency and the magnetic
field.

Some possible mechanisms of a selective emergence of the irregular
FS in the microwave phonon power spectra, the coexistence of
regular and irregular phaser spatial subsystems and the nature of
superslow transient processes (self-organized slowing-down) in an
autonomous phaser have been discussed.

The analogous phenomena of self-organized slowing-down have been
revealed recently by S.~D.~Makovetskiy in computer experiments with
a discrete autonomous excitable systems (DAES)
\cite{b35-SDM-cond-mat-0602345,b36-SDM-Master-Thesis-2006}, which
can be regarded as simplified models of phaser media.

One of the typical scenarios of DAES evolution is the emergence and
very slow (Zeno-like) development of complicated coherent
autostructures (e.~g., rotating spiral waves) which may spatially
coexist with irregular ones \cite{b37-SDM-cond-mat-0410460}. The
numerically revealed phenomenon of the bottlenecked evolution of
DAES may have the direct relation to the slowing-down of transient
processes that were observed experimentally in this work for an
autonomous phaser.

\clearpage


\begin{center}

{\textbf{FIGURE CAPTIONS}}

\vspace{10pt}

to the paper of D.~N.~Makovetskii

\vspace{10pt}

``Self-Organized Bottleneck and Coexistence of Incongruous States
in a Microwave Phonon Laser (Phaser)''

\vspace{10pt}

(figures see as separate PNG- and JPG-files)

\end{center}

\textbf{Fig.~1:} Microwave power spectrum of longitudinal phonon SE
at $\Delta_P = \Delta_H = 0$ in the ruby phaser. The frequency
interval between neighboring CS modes of phonon SE amounts to
310~kHz. The cryostat temperature is $\theta = 1.7$~K.

\vspace{10pt}

\textbf{Fig.~2:} Emergence of a regular FS in the power spectra of
an autonomous phaser at $P = 4$~mW, $\Delta_P = 8.8$~MHz, $\theta =
1.8$~K, and for various $\Delta_H$. Left oscillogram: $\Delta_H =
0$ (the FS is absent). Middle oscillogram: $\Delta_H = $~1,5~Oe
(the two-component FS). Right oscillogram: $\Delta_H = $~3,5~Oe
(the three-component FS). The range of the frequency sweeping along
the horisontal axis is about 6~kHz for each oscillogram.

\vspace{10pt}

\textbf{Fig.~3:} Chaotic FS of the phonon generation at $\theta =
1.8$~K, $P = 4$~mW, $\Delta_P = 8.8$~MHz, and $\Delta_H = 15$~Oe.
The range of the frequency sweeping along the horisontal axis is
about 18~kHz.

\vspace{10pt}

\textbf{Fig.~4:} The same as in Fig.~3, but at reduced pump power
$P = 0.9$~mW.



\end{document}